\documentclass[
 superscriptaddress,
reprint,
 bibnotes,
 showkeys,
 amsmath,amssymb,
 aps,
]{revtex4-1}
\pdfoutput=1

\usepackage{graphicx}
\usepackage{amsmath}
\usepackage{dcolumn}
\usepackage{bm}
\usepackage{braket} 
\usepackage{bbold}
\usepackage{xspace}

\usepackage[
 colorlinks=true,
 urlcolor=blue,
 citecolor=blue,
 linkcolor=blue,
 bookmarks=false,
 pdfstartview={FitH},
]{hyperref}

\usepackage{bm}
\usepackage{graphicx}
\usepackage{color}
\usepackage{xspace}
\definecolor{dgreen}{rgb}{0,0.4,0}
\definecolor{dpink}{rgb}{0.8,0,0.8}

\begin{document}

\title{ARPES observation of Mn-pnictide hybridization and negligible band structure renormalization in BaMn$_2$As$_2$ and BaMn$_2$Sb$_2$}

\author{W.-L. Zhang}
\affiliation{Beijing National Laboratory for Condensed Matter Physics, and Institute of Physics, Chinese Academy of Sciences, Beijing 100190, China}
\author{P. Richard}\email{p.richard@iphy.ac.cn}
\affiliation{Beijing National Laboratory for Condensed Matter Physics, and Institute of Physics, Chinese Academy of Sciences, Beijing 100190, China}
\affiliation{School of Physical Sciences, University of Chinese Academy of Sciences, Beijing 100190, China}
\affiliation{Collaborative Innovation Center of Quantum Matter, Beijing, China}
\author{A. van Roekeghem}
\affiliation{Beijing National Laboratory for Condensed Matter Physics, and Institute of Physics, Chinese Academy of Sciences, Beijing 100190, China}
\affiliation{Centre de Physique Th\'{e}orique, Ecole Polytechnique, Universit\'{e} Paris-Saclay, 91128 Palaiseau, France}
\affiliation{CEA, LITEN, 17 Rue des Martyrs, 38054 Grenoble, France}
\author{S.-M. Nie}
\affiliation{Beijing National Laboratory for Condensed Matter Physics, and Institute of Physics, Chinese Academy of Sciences, Beijing 100190, China}
\author{N. Xu}
\affiliation{Beijing National Laboratory for Condensed Matter Physics, and Institute of Physics, Chinese Academy of Sciences, Beijing 100190, China}
\affiliation{Paul Scherrer Institut, Swiss Light Source, CH-5232 Villigen PSI, Switzerland}
\author{P.~Zhang}
\affiliation{Beijing National Laboratory for Condensed Matter Physics, and Institute of Physics, Chinese Academy of Sciences, Beijing 100190, China}
\affiliation{ISSP, University of Tokyo, Kashiwa, Chiba 277-8581, Japan}
\author{H. Miao}
\affiliation{Beijing National Laboratory for Condensed Matter Physics, and Institute of Physics, Chinese Academy of Sciences, Beijing 100190, China}
\affiliation{Condensed Matter Physics and Materials Science Department, Brookhaven National Laboratory, Upton, New York 11973, USA}
\author{S.-F. Wu}
\affiliation{Beijing National Laboratory for Condensed Matter Physics, and Institute of Physics, Chinese Academy of Sciences, Beijing 100190, China}
\author{J.-X.~Yin}
\affiliation{Beijing National Laboratory for Condensed Matter Physics, and Institute of Physics, Chinese Academy of Sciences, Beijing 100190, China}
\author{B.~B.~Fu}
\affiliation{Beijing National Laboratory for Condensed Matter Physics, and Institute of Physics, Chinese Academy of Sciences, Beijing 100190, China}
\author{L.-Y. Kong}
\affiliation{Beijing National Laboratory for Condensed Matter Physics, and Institute of Physics, Chinese Academy of Sciences, Beijing 100190, China}
\author{T. Qian}
\affiliation{Beijing National Laboratory for Condensed Matter Physics, and Institute of Physics, Chinese Academy of Sciences, Beijing 100190, China}
\affiliation{Collaborative Innovation Center of Quantum Matter, Beijing, China}
\author{Z.-J. Wang}
\affiliation{Beijing National Laboratory for Condensed Matter Physics, and Institute of Physics, Chinese Academy of Sciences, Beijing 100190, China}
\author{Z. Fang}
\affiliation{Beijing National Laboratory for Condensed Matter Physics, and Institute of Physics, Chinese Academy of Sciences, Beijing 100190, China}
\affiliation{Collaborative Innovation Center of Quantum Matter, Beijing, China}
\author{A. S. Sefat}
\affiliation{Materials Science and Technology Division, Oak Ridge National Laboratory, Oak Ridge, Tennessee 37831-6114, USA}
\author{S. Biermann}
\affiliation{Centre de Physique Th\'{e}orique, Ecole Polytechnique, Universit\'{e} Paris-Saclay, 91128 Palaiseau, France}
\affiliation{Coll\`{e}ge de France, 11, place Marcelin-Berthelot, 75005 Paris, France}
\affiliation{European Theoretical Synchrotron Facility, Europe, France}
\author{H. Ding}\email{dingh@iphy.ac.cn}
\affiliation{Beijing National Laboratory for Condensed Matter Physics, and Institute of Physics, Chinese Academy of Sciences, Beijing 100190, China}
\affiliation{School of Physical Sciences, University of Chinese Academy of Sciences, Beijing 100190, China}
\affiliation{Collaborative Innovation Center of Quantum Matter, Beijing, China}

\date{\today}

\begin{abstract}

We performed an angle-resolved photoemission spectroscopy study of  BaMn$_2$As$_2$ and BaMn$_2$Sb$_2$, which are isostructural to the parent compound BaFe$_2$As$_2$ of the 122 family of ferropnictide superconductors. We show the existence of a strongly $k_z$-dependent band gap with a minimum at the Brillouin zone center, in agreement with their semiconducting properties. 
Despite the half-filling of the electronic 3$d$ shell, we show that the band structure in these materials is almost not renormalized from the Kohn-Sham bands of density functional theory. Our photon energy dependent study provides evidence for Mn-pnictide hybridization, which may play a role in tuning the electronic correlations in these compounds.
\end{abstract}

\pacs{74.70.Xa, 74.25.Jb, 79.60.-i, 71.18.+y}


\maketitle

\section{Introduction}

Angle-resolved photoemission spectroscopy (ARPES) experiments on Fe-based superconductors indicate non-negligible band renormalization due to electronic correlations \cite{RichardRoPP2011,Ambroise_CR_Physique2016}. The key role attributed to a significant Hund's rule coupling in these materials and in their isostructural nonferropnictide counterparts in tuning the electronic correlations depends strongly on the 3$d$ electronic shell filling \cite{HauleNJP11,AichhornPRB82,LiebschPRB82,IshidaPRB81,AGeorges_AnnuRev_2013}. The electronic correlations are expected to reach a maximum at half-filling (Mn 3$d^5$) and to decrease away from that configuration. This scenario is consistent with the observation by ARPES of reduced electronic correlations in BaCo$_2$As$_2$ as compared to BaFe$_2$As$_2$ \cite{Nan_XuPRX3}, which is inferred from a smaller renormalization of the electronic band structure in the former case. However, how the electronic correlations, expected to be even larger for a half-filled Mn 3$d^5$ shell, affect the band renormalization in BaMn$_2$As$_2$, has not been addressed experimentally.

Here we report an ARPES study of BaMn$_2$As$_2$ and BaMn$_2$Sb$_2$, which have the same crystal structure as the ferropnictide parent compound BaFe$_2$As$_2$. We find a $k_z$-dependent band gap at the Brillouin zone (BZ) center compatible with the semiconducting behavior of this material \cite{J_AnPRB79}. 
Despite the half-filling of the Mn 3$d$-shell, we find that the electronic band structure of BaMn$_2$As$_2$ and BaMn$_2$Sb$_2$ is basically unrenormalized compared to Kohn-Sham bands of density functional theory. 
While the antiferromagnetic order present in these compounds is expected to
milden correlation-induced band renormalizations, this effect is stronger
than in antiferromagnetic BaFe$_2$As$_2$.
Interestingly, our photon energy ($h\nu$) measurements show Mn-pnictide hybridization for the electronic states near the Fermi level energy ($E_F$). Our study suggests that the hybridization between the 3$d$ transition metal atoms and the pnictide atoms cannot be neglected in evaluating the strength of the electronic correlations in the 122 materials.

\section{Experiment and calculation method}

The BaMn$_2$As$_2$ and BaMn$_2$Sb$_2$ single crystals used in this study were grown by the flux method \cite{J_AnPRB79}. ARPES experiments were performed at 25 K at beamlines PGM and APPLE-PGM of the Synchrotron Radiation Center (WI), with a VG-Scienta R4000 analyzer and a VG-Scienta SES 200 analyzer, respectively. The energy and angular resolutions were set at 15-30 meV and 0.2$^\circ$, respectively. The synchrotron data were recorded with $\sigma$ polarized light tracking electronic states that are odd with respect to the photoemission plane \cite{XP_WangPRB85}. We point out that the configuration used is not pure and that the incident light contains a non-negligible component of the potential vector along the direction perpendicular to the sample, which probes orbitals extended along the $z$ axis, such as $d_{z^2}$ and $p_z$.  Complementary measurements at 30 K and 300 K were performed at the Institute of Physics, Chinese Academy of Sciences, using the He $\alpha$I resonance line of a helium discharge lamp ($h\nu=21.218$ eV). The light from the He discharge lamp is unpolarized, except for a component of the potential vector along perpendicular to the sample surface. The samples were cleaved \textit{in situ} and measured in a vacuum better than 5$\times10^{-11}$ Torr. The $E_F$ of the samples was referenced to that of a freshly evaporated Ag or Au polycrystalline film. Following our previous works, we display the experimental results using the 1 Mn/unit cell notations for the BZ, with $a$ corresponding to the distance between Mn first neighbors and $c^{\prime}=c/2$ corresponding to the distance between two Mn planes. The first-principles calculations of the electronic band structures are performed in the G-type antiferromagnetic order, in which the unit cell contains 2 Mn atoms, as in the paramagnetic case. These calculations were done using the full-potential linearized-augmented plane-wave (FP-LAPW) method implemented in the WIEN2K package for 
the experimental crystal structures~\cite{LAPW} and
assuming the G-type antiferromagnetic order observed
experimentally. 
The exchange-correlation potential was treated using the generalized gradient approximation (GGA) 
by Perdew, Burke and Ernzerhof (PBE)~\cite{Perdew_PRL77}. The radius of the muffin-tin sphere $R_{MT}$ were 2.5, 2.5 and 2.26 Bohr for Ba, Mn and As, respectively. A $10\times 10\times10$ $k$-point mesh has been utilized in the self-consistent calculations. The truncation of the modulus of the reciprocal lattice vector $K_{max}$, which was used for the expansion of the wave functions in the interstitial regions, was set to $R_{MT}\times K_{max} = 7$.

\section{Results and discussion}

\begin{figure}[!t]
\begin{center}
 \includegraphics[width=8.5cm]{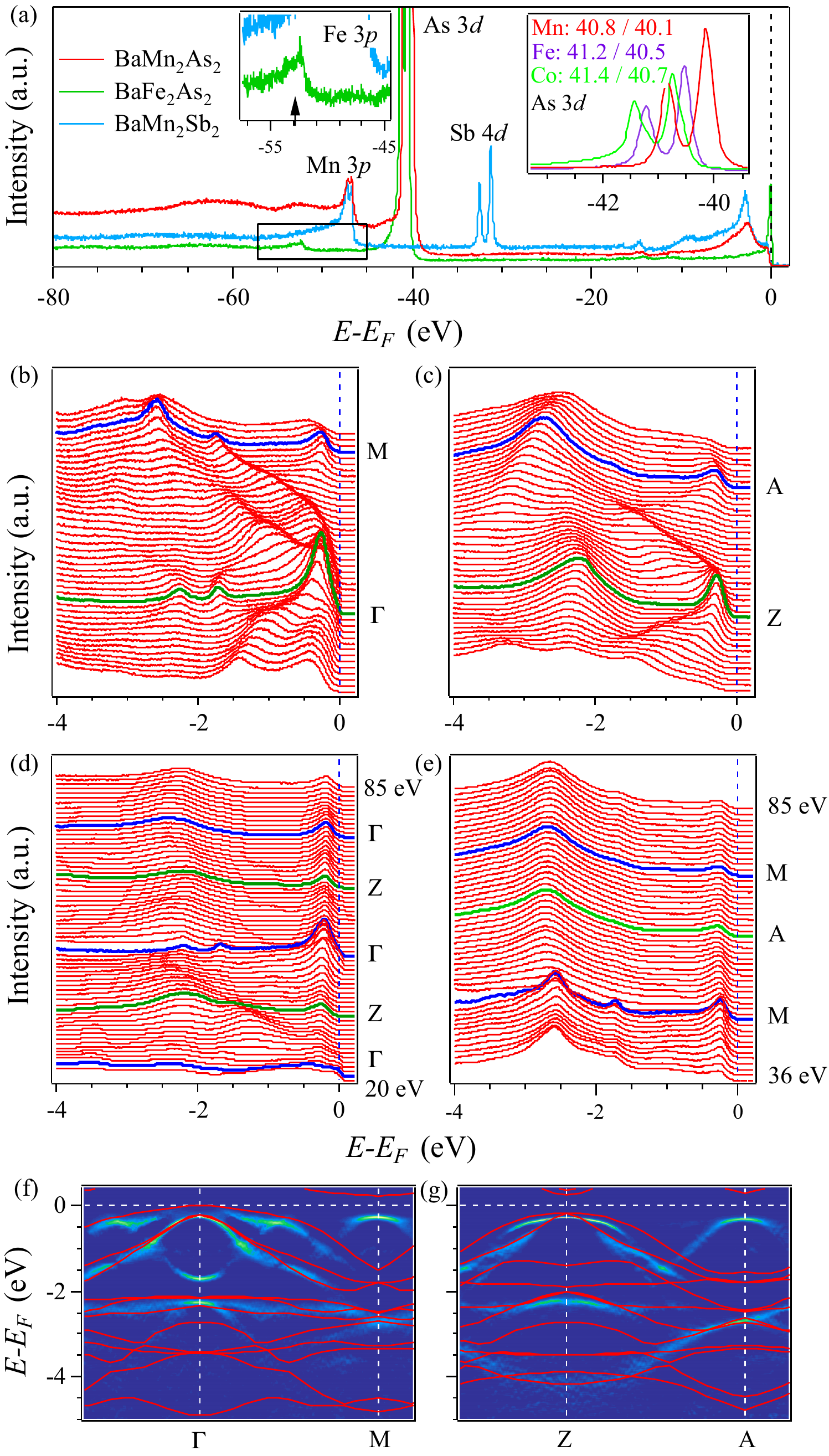}
 \end{center}
 \caption{\label{Fig: A}(color online) (a) Core level spectra of BaMn$_2$As$_2$, BaFe$_2$As$_2$ (green) and BaMn$_2$Sb$_2$ (blue). The inset shows a zoom on the As $3d$ levels of BaMn$_2$As$_2$ (red), BaFe$_2$As$_2$ (green) and BaCo$_2$As$_2$ (purple). (b) and (c) EDCs of BaMn$_2$As$_2$ measured at 25 K along the $\Gamma$-M ($k_z = 0$) and Z-A ($k_z = \pi$) high-symmetry lines, respectively. (d) $h\nu$ dependence of the normal emission EDCs in BaMn$_2$As$_2$. (e) $h\nu$ dependence of the EDCs in BaMn$_2$As$_2$ along M-A. (f) and (g) Intensity plots of 2D-curvature \cite{P_Zhang_RSI2011} corresponding to the EDCs in (b) and (c), respectively. Non-renormalized GGA bands are overlapped for comparison.}
\end{figure}

\begin{figure}[!t]
\begin{center}
 \includegraphics[width=8.5cm]{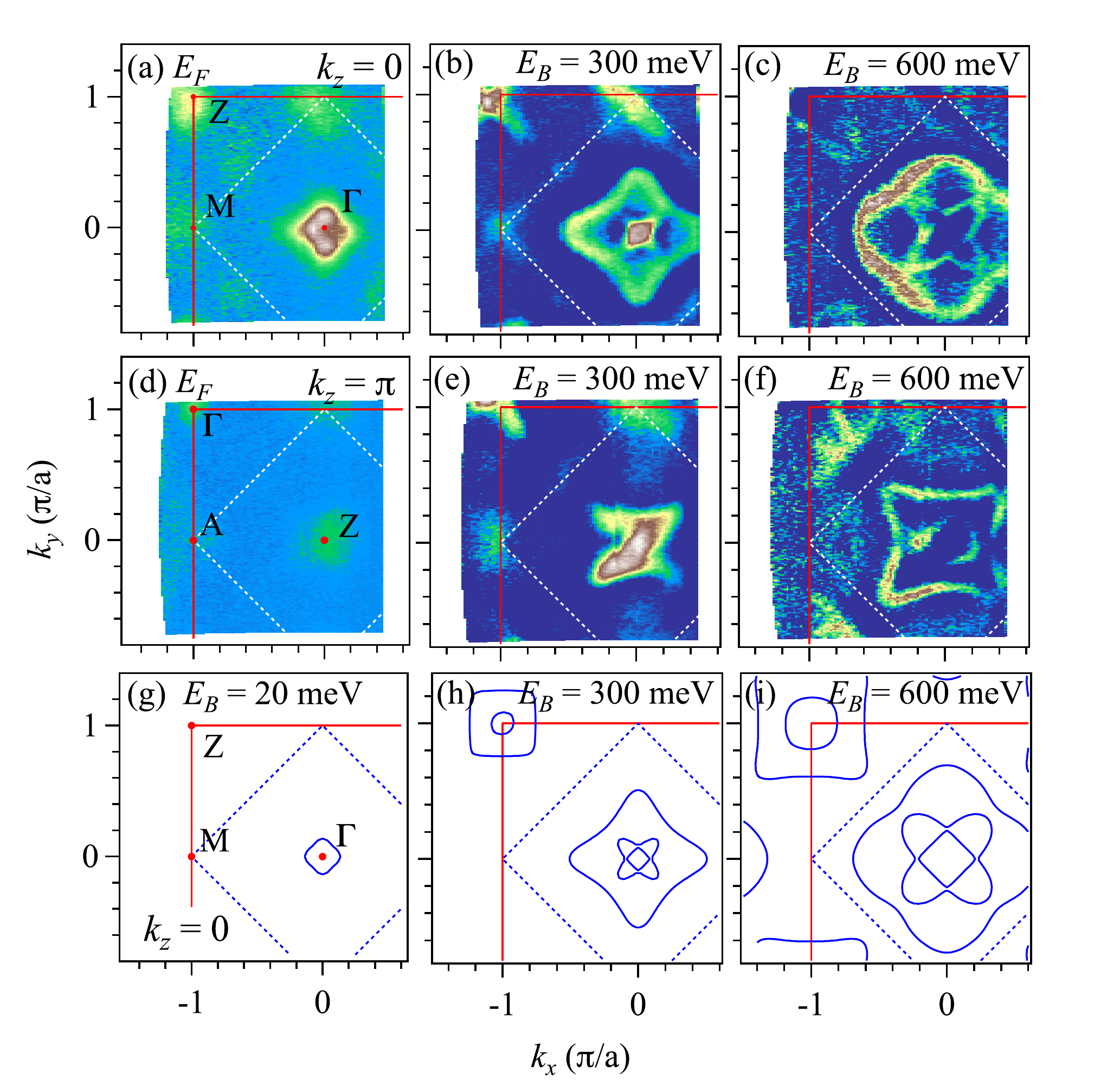}
 \end{center}
 \caption{\label{Fig: B}(color online) (a) ARPES intensity map at $E_F$ in the $k_z = 0$ plane of BaMn$_2$As$_2$. (b) and (c) 1D curvature energy contour maps at 300 meV and 600 meV below $E_F$ at $k_z = 0$ plane. (d)-(f) Same as (a)-(c) but in the $k_z=\pi$ plane. The energy integration window is 20 meV for all plots. (g)-(i) Density functional calculations in the $k_z=0$ plane at $E_B=20$ meV, $E_B=300$ meV and $E_B=600$ meV, respectively.}
\end{figure}

\begin{figure*}[!t]
 \begin{center}
 \includegraphics[width=18cm]{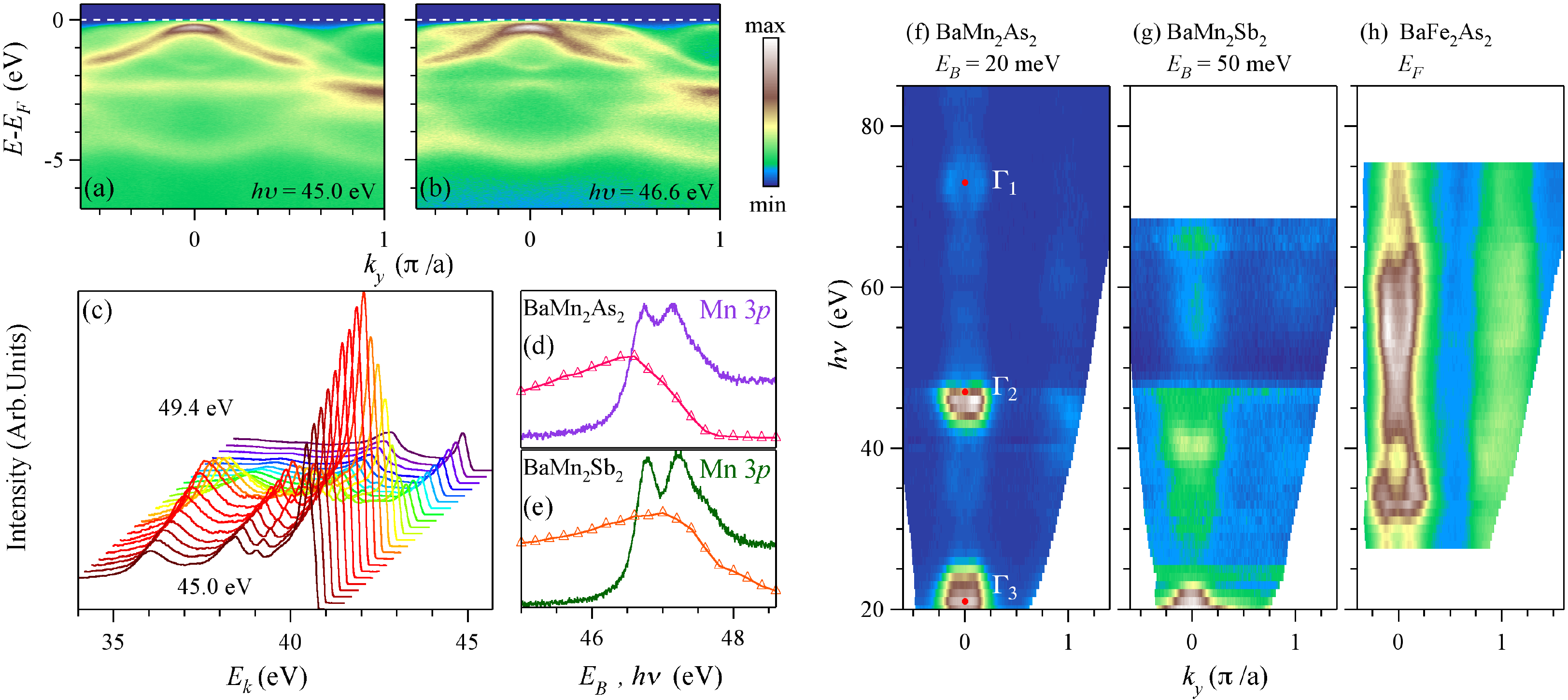}
 \end{center}
 \caption{\label{Fig: C}(color online) (a) and (b) ARPES intensity plots along $\Gamma-M$ recorded in BaMn$_2$As$_2$ at $h\nu=45.0$ and 46.6 eV, respectively. (c) Normal emission EDCs of BaMn$_2$As$_2$ as a function of $h\nu$ around the 46.6 eV resonance. (d) Integrated photoemission intensity $I$ (from $E_B=240$ meV to 260 meV) of the normal emission EDCs with respect to $h\nu$, and comparison with the Mn $3p$ core level spectrum of BaMn$_2$As$_2$. (e) Same as (d) but for BaMn$_2$Sb$_2$. (f)-(h) $h\nu$ dependence of the ARPES intensity along $\Gamma$-M in BaMn$_2$As$_2$ ($E_B=20$ meV), BaMn$_2$Sb$_2$ ($E_B=50$ meV) and BaFe$_2$As$_2$ ($E_F$), respectively.}
 \end{figure*}

The shallow core level spectra of BaMn$_2$As$_2$, BaFe$_2$As$_2$ and BaMn$_2$Sb$_2$, compared in  Fig.~\ref{Fig: A}(a), reveal their chemical composition. In contrast to the relatively broad and weak Fe 3$p$ peak observed at 52.7 eV in BaFe$_2$As$_2$, a well-defined double-peak structure corresponding to Mn 3$p$ electrons is observed around a binding energy ($E_B$) of 47 eV in both BaMn$_2$As$_2$ and BaMn$_2$Sb$_2$. While the Sb 4$d$ peaks detected at $E_B$ = 32.5 eV and 31.2 eV in BaMn$_2$Sb$_2$ cannot be compared directly to BaFe$_2$As$_2$, the shift in the position of the As 3$d$ peaks in BaTM$_2$As$_2$ (TM = Mn, Fe, Co) illustrates quite well the different fillings of the 3$d$ electron shells associated with different transition metals. As shown in the inset of Fig.~\ref{Fig: A}(a), the As 3$d$ peaks are downshifted by about 400 meV in BaMn$_2$As$_2$ as compared to BaFe$_2$As$_2$, which is consistent with a much smaller 3$d$ band filling. In contrast, a 200 meV upward shift attributed to electron doping was detected in BaCo$_2$As$_2$ \cite{Nan_XuPRX3}. We caution though that a simple rigid band shift is far from sufficient to describe the spectra, especially near $E_F$. In particular, a peak in the electronic density-of-states of BaMn$_2$As$_2$ and BaMn$_2$Sb$_2$ that develops around $E_B$ = 3 eV has no equivalent in BaFe$_2$As$_2$ at higher $E_B$, suggesting a rearrangement of the electronic states.

We performed ARPES experiments over a wide $h\nu$ range to investigate the near-$E_F$ electronic band structure in the 3D momentum space. In Figs.~\ref{Fig: A}(b) and \ref{Fig: A}(c) we plot the energy distribution curves (EDCs) along the $\Gamma$-M ($k_z=0$) and Z-A ($k_z=\pi$) high-symmetry lines, respectively. The corresponding intensity plots of 2D-curvature \cite{P_Zhang_RSI2011} are given in Figs.~\ref{Fig: A}(f) and \ref{Fig: A}(g), respectively, along with the electronic band dispersions derived from our GGA calculations. The 2D-curvature is an improved version of the laplacian to track dispersive features in image plots, which is based on the Gauss theory of curvature. We observe two hole bands at the BZ center, none of them crossing $E_F$, as expected from the antiferromagnetic ground state and semiconducting behavior of BaMn$_2$As$_2$ \cite{J_AnPRB79,SQ_XiaEJIC2008,HF_WangJAC477}. While the inner hole band is gapped by about 200 meV below $E_F$ for both $k_z$ values, the band gap of the outer one varies from nearly 93 to 300 meV between $\Gamma$ and Z. However, due to band broadness, the tail of spectral intensity extends to lower energies, and the leading edge of the EDC at $\Gamma$ is as low as 10 meV, which clarifies why the system has a metallic-like resistivity above 100 K \cite{J_AnPRB79} despite the large gap reported earlier from ARPES at a single $h\nu$ value \cite{PandeyPRL108}. The same conclusion is drawn from the $h\nu$ dependence of the normal emission EDCs, which are displayed in Fig.~\ref{Fig: A}(d). 

The dispersion of the two $\Gamma$-centered hole bands is rather well reproduced by our non-renormalized GGA calculations. In fact, a simple downward shift of 50 meV is sufficient to capture most of the ARPES dispersive features down to 5 eV below $E_F$. Despite this remarkable agreement, some bands cannot be explained by the calculations. In particular, we can distinguish an additional hole band at the M and A points, which is quite similar to the inner $\Gamma$-centered hole band. Our experimental results also reveal one band reaching its bottom near -1.7 eV at the $\Gamma$ point. Interestingly, a similar band appears clearly in the EDCs of Fig. \ref{Fig: A}(b) at the M point, thus suggesting a $\Gamma$-M band folding, as reported earlier \cite{PandeyPRL108}. We note that the band folding is also noticeable in the $h\nu$ dependence of the EDCs along M-A displayed in Fig.~\ref{Fig: A}(e), which show a peak at the same energy as along the $\Gamma$-Z direction.

To investigate further the band folding, we show in Fig. \ref{Fig: B} the maps of ARPES intensity at $E_F$ as well as the intensity maps of 1D curvature along $k_y$ at 300 meV and 600 meV below $E_F$. Despite the absence of a real Fermi surface, we observe small residual photoemission intensity at $\Gamma$ [Fig. \ref{Fig: B}(a)] and Z [Fig. \ref{Fig: B}(d)] due to the tail of the spectral weight associated with the $\Gamma$-centered hole bands. In addition, we detect unexpected intensity at the M and A points, reinforcing our assumption of a $\Gamma$-M band folding. At 300 meV below $E_F$ the intensity map of curvature reveals a pattern at the $\Gamma$ point formed by a diamond-like contour enclosing a X-shape feature. While we do not detect any trace of the diamond-like constant energy contour at the M point, the X-shape feature is clearly observed. Our calculations of constant energy maps, displayed in Figs. \ref{Fig: B}(g)-\ref{Fig: B}(i), show patterns at the $\Gamma$ and Z points that are consistent with the experimental observations. However, nothing is predicted at the M point, in contrast to the ARPES data.

The reason for the ARPES observation of such band folding is unclear. Neutron diffraction experiments on BaMn$_2$As$_2$ suggest a G-type AFM alignment with Mn moments aligned along the $c$-axis \cite{Y_SinghPRB80}. Unlike the collinear magnetic order usually observed in the 122 ferropnictides \cite{Huang_PRL101, J_Zhao_PRL2008}, which induces a $\Gamma$-M band folding \cite{Richard_PRL2010,Scalapino_PRB2010}, no in-plane folding is expected for the G-type AFM structure in the 122 materials, whatever the in-plane or out-of-plane alignment of the Mn moments. Our results suggest that there is a $\sqrt{2}\times\sqrt{2}$ doubling of the primitive cell, if not in the bulk at least at the surface, caused by either a magnetic or crystal structure distortion. Although it cannot induce the band folding observed in BaMn$_2$As$_2$, a weak in-plane ferromagnetic component was reported in hole-doped Ba$_{1-x}$K$_x$Mn$_2$As$_2$ \cite{JK_BaoPRB85,PandeyPRL111}. Although a theoretical study attributes the in-plane ferromagnetic component to a canting of the Mn moments \cite{Glasbrenner2013}, recent X-ray magnetic circular dichroism experiments rather suggest that the ferromagnetic component resides on the As 2$p$ orbital \cite{Ueland_PRL114}. In either cases, the magnetic structure in BaMn$_2$As$_2$ is certainly prone to a distortion as compared to the simple G-type AFM of the Mn alone, which could possibly induce a band folding. Alternatively, a $\sqrt{2}\times\sqrt{2}$ reconstruction 
at the surface could also lead to the observed band folding~\cite{Massee_PRB80,Nascimento_PRL103}. We note that neither the Sb $4d$ nor the As $3d$ core levels exhibit anomalies similar to that reported in the EuFe$_2$As$_{2-x}$P$_x$ and associated to a surface state~\cite{Richard_JPCM26}.

Except for the band folding described above, the relatively good agreement between the GGA calculations and the ARPES data is in sharp apparent contrast with the common expectation that the half-filled 3$d$ band of BaMn$_2$As$_2$ would lead to electronic states more renormalized than for the 3$d^6$ configuration of the ferropnictides \cite{YX_YaoPRB84}, for which overall renormalization factors of about 2-5 are typically reported by ARPES for the Fe $3d$ states, regardless of antiferromagnetic ordering \cite{RichardRoPP2011}. In order to understand this effect, we investigated the elemental composition of the electronic states near $E_F$ by performing photoemission experiments over a wide $h\nu$ range. Strong variation of photoemission intensity is observed, in particular in the $h\nu=45-49$ eV range. For example, a strong contrast of intensity is observed between ARPES cuts recorded with 45 eV and 46.6 eV photons, as shown in Figs. \ref{Fig: C}(a) and \ref{Fig: C}(b), respectively. In particular, the outmost hole band around the $\Gamma$ point and the band that bottoms at 4.5 eV are strongly affected. Fig. \ref{Fig: C}(c) illustrates the evolution of the photoemission intensity of the normal emission EDCs within the $h\nu = 45-49.4$ eV range. A resonance, followed by a strong decrease in the overall intensity, is observed around 46.6 eV. As shown in Fig.~\ref{Fig: C}(d), the profile $I$ of photoemission intensity at $\Gamma$ as a function of $h\nu$ for the 240 meV - 260 meV $E_B$ range is in good coincidence with the Mn $3p$ core levels displayed as a function of $E_B$. As shown in Fig.~\ref{Fig: C}(e), a similar phenomenon is observed in BaMn$_2$Sb$_2$ at a slightly higher resonance energy ($h\nu=47$ eV). Although it contrasts with the anti-resonance profile reported in Ba$_{0.6}$K$_{0.4}$Fe$_2$As$_2$ at 53 eV and coinciding with the Fe $3p$ absorption edge \cite{Ding_JPCM2011}, this result confirms that the near-$E_F$ electronic states in BaMn$_2$As$_2$ and BaMn$_2$Sb$_2$ derive for a large part from Mn $3d$ orbitals. 

\begin{figure*}[!t]
\begin{center}
 \includegraphics[width=15cm]{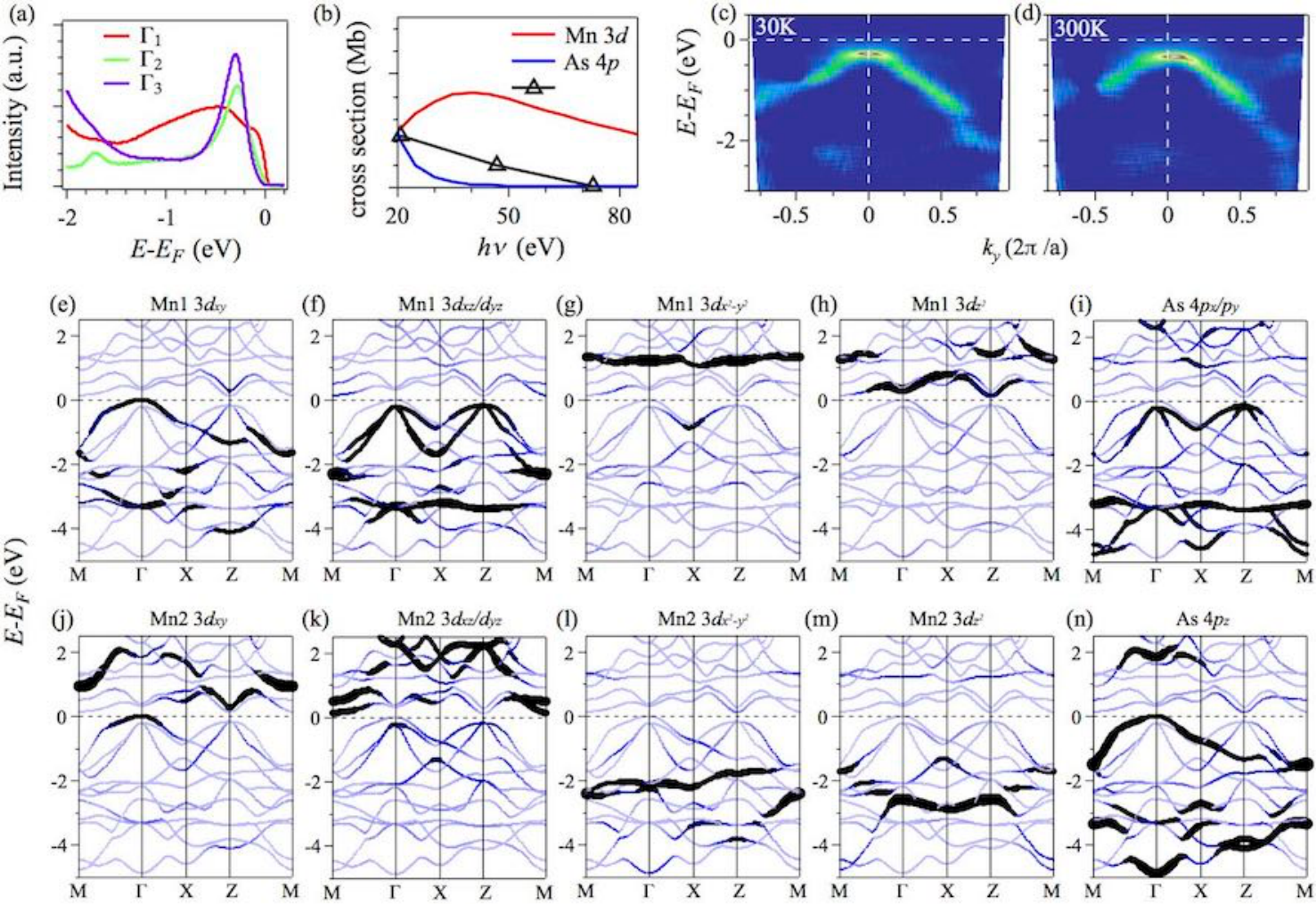}
\end{center}
 \caption{\label{Fig: D}(color online) (a) Normal emission EDCs at different $\Gamma$ points [$\Gamma_1(h\nu=21$~eV), $\Gamma_2(h\nu=47$~eV) and $\Gamma_3(h\nu=73$~eV)] of BaMn$_2$As$_2$.  (b) Integrated intensity (20~meV below E$_F$) of the normal emission EDCs and the calculated cross section of photoionization for the Mn 3\textit{d} and As 4\textit{p} orbitals \cite{Yeh_Lindau1985}. (c) and (d) Intensity plots of 2D curvature of cuts recorded in BaMn$_2$As$_2$ along $\Gamma$-M at 30~K and 300~K, respectively,  using the He $\alpha$I resonance line of a helium discharge lamp. (e)- (n) GGA calculations of BaMn$_2$As$_2$ in the antiferromagnetic state, along high-symmetry lines. Mn1 and Mn2 refer to Mn sites with spin up (majority state) and spin down (minority state), respectively. The size of the markers indicate the orbital projection on the states indicated on the top of each panel. For a better identification of the As $4p$ character, the size of the markers for those states has been chosen 5 times larger than the one for the Mn $3d$ states. }
\end{figure*}

Careful examination of the $h\nu$ dependence data reveals another photoemission behavior. We plot in Figs.~\ref{Fig: C}(f)-\ref{Fig: C}(h) the $h\nu$ dependent intensity map in the $k_x=0$ plane. The 20 meV energy integration windows are centered at $E_B=20$ meV, 50 meV and 0 in BaMn$_2$As$_2$, BaMn$_2$Sb$_2$ and BaFe$_2$As$_2$, respectively. Interestingly, stronger spots of intensity in Figs. \ref{Fig: C}(f) and \ref{Fig: C}(g) are observed for $h\nu$ values coinciding to the $\Gamma$ point, but this intensity varies from one $\Gamma$ position to the next. In Fig. \ref{Fig: D}(a) we compare the EDCs of BaMn$_2$As$_2$ at 3 $\Gamma$ points [$\Gamma_1(h\nu=21$ eV), $\Gamma_2(h\nu=47$ eV) and $\Gamma_3(h\nu=73$ eV)], and we plot in Fig. \ref{Fig: D}(b) the curve of the integrated value of the near-$E_F$ intensity as a function of $h\nu$. While the calculated photoemission cross section of the Mn $3d$ electrons \cite{Yeh_Lindau1985}, also displayed in Fig. \ref{Fig: D}(b), is the smallest for the lowest photon energies and shows a maximum around 40 eV, the experimental intensity decreases monotonically with $h\nu$, which is more consistent with As $4p$ states. However, this experimental decrease is slower than expected for the As $4p$ states, most likely due to the hybridization with the Mn $3d$ states. As reported in a previous theoretical study \cite{J_AnPRB79}, the Mn 3$d$ and As 4$p$ orbital projections obtained from our GGA calculations, displayed in Figs.~\ref{Fig: D}(e)-~\ref{Fig: D}(n), suggest a certain level of Mn-pnictide hybridization in these systems. This is notably true for the $d_{xy}$ band that approaches $E_F$ at the zone center, which hybridizes with the $p_z$ orbital. We note that the enhanced $k_z$ dispersion of the $d_{xy}$ band observed in BaMn$_2$As$_2$ as compared to BaFe$_2$As$_2$ is consistent with its hybridization with the $p_z$ orbital. We note that the bands that are observed are consistent with the $\sigma+A_z$ polarization configuration of our experimental setup.

We now discuss the absence of strong band renormalization in BaMn$_2$As$_2$ and BaMn$_2$Sb$_2$. A recent study shows that density functional theory (DFT) calculations, which do not include the effect of the electronic correlations, give results that are quite different from the spectral function calculated by DFT+dynamical mean field theory (DFT+DMFT) in the paramagnetic state of LaOMnAs \cite{Zingl_PRB94}, thus expressing the effect of the electronic correlations. It is also argued that the proximity to a Mott transition in LaOMnAs and BaMn$_2$As$_2$ is responsible for their high N\'{e}el temperatures  $T_N$'s  \cite{Zingl_PRB94}. For example, BaMn$_2$As$_2$ exhibits G-type antiferromagnetism with a N\'{e}el temperature as high as $T_N= 625$ K and a large ordered moment ($\mu=3.9\mu_B$) \cite{Y_SinghPRB80}. Interestingly, the calculated DFT+DMFT spectral function of LaOMnAs gains coherence in the antiferromagnetic state, with the DFT bands overlapping pretty well on the DFT+DMFT calculations. 
This supports the idea that the first ingredient being responsible for the negligible band renormalization is indeed the antiferromagnetic order of the compounds. This observation matches what is generally known from DFT calculations: one of the most important failure of DFT is the inability of describing local moment behavior in paramagnetic phases. Those are either described as non-magnetic or one has to artificially introduce magnetic order. In the presence of magnetic order, many of the consequences of the existence of magnetic moments can in fact be correctly described within a one-particle picture, rationalizing why magnetic calculations often give a more appropriate picture of certain physical properties. However, the situation in BaMn$_2$As$_2$ goes beyond this relatively simple observation. Indeed, the case of BaMn$_2$As$_2$ contrasts with the non-negligible band renormalization found in the antiferromagnetic state of metallic BaFe$_2$As$_2$, well below $T_N$ \cite{Richard_PRL2010}.
One may speculate that the origin of this difference lies precisely in the
half-filled shell: Hund's coupling induces a high-spin configuration, which
is maximally efficient in suppressing charge and spin fluctuations, restoring
an unrenormalized band picture.
This is consistent with findings within DFT + DMFT calculations that indicate 
that Hund's coupling ($J_H$) is responsible for the insulating nature of BaMn$_2$As$_2$ \cite{McNally_PRB92}, and that the (metallic) band structure 
calculated without Hund's coupling is significantly more renormalized than 
in the calculation that includes a finite $J_H$.

Transport measurements show an insulating-like behavior for BaMn$_2$As$_2$ 
at the temperature for which most of our ARPES data were recorded. 
This observation directly documents the absence of charge fluctuations 
(which goes hand in hand with a suppression of spin fluctuations due to the $d^5$ high spin configuration). Increasing the number of itinerant carriers
could in this picture be expected to reintroduce band renormalisation.
To check this possibility, we compare ARPES data obtained at 30~K and 
300 K in Figs.~\ref{Fig: D}(c)~-~\ref{Fig: D}(d). Although smaller than 
$T_N$, the latter temperature is much higher than 100~K, temperature 
around which resistivity switches from insulating-like to metallic-like 
\cite{J_AnPRB79}. Besides thermal broadening, the results are basically 
identical, and no change in the bandwidth is observed at 300~K as 
compared to 30~K, suggesting that the carrier density might still be too 
low.

\section{Summary}

In summary, we have used ARPES to characterize the electronic band structure of BaMn$_2$As$_2$ and BaMn$_2$Sb$_2$. We have shown that their electronic structure is consistent with their semiconducting properties. We observed a strongly $k_z$-dependent band gap with a minimum of 93 meV at $k_z=0$. Our data provide an experimental proof of a Mn-pnictide hybridization in these compounds. One direct consequence is 
a larger spatial extension of the Mn $3d$ states, which should thus lead to a decrease in the electronic correlations. Our results suggest that the network of the transition metal atoms in the ferropnictides and related nonferropnictides cannot be viewed as independent in evaluating the strength of the electronic correlations in these materials. 

 \section*{Acknowledgement}

We acknowledge G. Kotliar for useful discussions. This work was supported by grants from MOST (2011CBA001000, 2011CBA00102, 2012CB821403, 2013CB921703 and 2015CB921301) and NSFC (11004232, 11034011/A0402, 11234014, 11274362 and 11674371) from China
and a Consolidator Grant of the European Research Council (project number
617196). This work is based in part on research conducted at the Synchrotron Radiation Center, which was primarily funded by the University of Wisconsin-Madison with supplemental support from facility Users and the University of Wisconsin-Milwaukee. The work at ORNL was supported by the Department of Energy, Basic Energy Sciences, Materials Sciences and Engineering Division.

\bibliography{BMA_biblio.bib}

\end{document}